# Spectral properties of interacting magnetoelectric particles

E.O. Kamenetskii

Ben-Gurion University of the Negev, Beer Sheva 84105, Israel

August 7, 2008

**Abstract**

The linear magnetoelectric (ME) effect provides a special route for linking magnetic and electric properties. In microwaves, a local ME effect appears due to the dynamical symmetry breakings of magnetic-dipolar modes (MDMs) in a ferrite disk particle. The fact that for MDMs in a ferrite disk one has evident both classical and quantum-like attributes, puts special demands on the methods used for study of interacting ME particles. A proper model for coupled particles should be based on the spectral characteristics of MDM oscillations and an analysis of the overlap integrals for interacting eigen oscillating ME elements. In this paper, we present theoretical studies of spectral properties of literally coupled of MDM ME disks. We show that there exists the "exchange" mechanism of interaction between the particles, which is distinctive from the magnetostatic interaction between magnetic dipoles. The spectral method proposed in this paper may further the development of a theory of ME "molecules" and realization of local ME composites.

## I. INTRODUCTION

The symmetry relationships between the electric polarization and the magnetization make questionable an idea of simple combination of two (electric and magnetic) dipoles to realize local magnetoelectric (ME) particles. The electric polarization is parity-odd and time-reversal-even. At the same time, the magnetization is parity-even and time-reversal-odd [1]. If one supposes that he has created an "artificial atom" with the local cross-polarization effect one, certainly, should demonstrate a special ME field in the near-field region. It means that using a gedankenexperiment with two quasistatic, electric and magnetic, point probes for the ME near-field characterization, one should observe not only an electrostatic-potential distribution (because of the electric polarization) and not only a magnetostatic-potential distribution (because of the magnetic polarization), one also should observe a special cross-potential term (because of the cross-polarization effect). This fact contradicts to classical electrodynamics. One cannot consider (classical electrodynamically) two coupled electric and magnetic dipoles – the ME particles – as local (near-field) sources of the electromagnetic field [1]. So in a presupposition that an "artificial atom" with the near-field cross-polarization effect is really created, one has to show that in this particle there are special internal dynamical motion processes different from the classical motion processes [2]. Numerous classical models of so called bianisotropic particles proposed in literature [3] do not provide the reader with real physics of ME coupling between the microscopic electric and magnetic currents.

Physically, there are different characterizations of magnetoelectricity, or ME effect. For example, one can characterize the ME effect as the appearance of an electric field in certain substances, when they are subjected to a static magnetic field. Another characterization of the ME effect is related to linear coupling between magnetization and polarization in solid-state structures. Natural magnetoelectric crystals are solid-state structures with linear coupling between magnetization and polarization. Physics of the ME effect in crystals becomes evident not from a pure classical basis. In different physical problems, ME coupling is due to symmetry breaking phenomena. In crystals and molecular systems, magnetoelectricity takes place when space inversion is locally broken [4]. If the ME effect exists, the interaction between electrons and elementary magnetic cells appears in such a way that the resulting local polarization and magnetization break the local relativistic crystalline symmetry. In natural crystals, ME properties are evident or at very low frequencies, or in an optical region. Recently, the microwave ME effect was demonstrated in layered ferrite/piezoelectric structures [5].

Natural magnetoelectric crystals and layered ferrite/piezoelectric structures are not materials composed by small structural elements – local ME particles. In a proposition of particulate ME composites one may suppose that the unified ME fields originated from a point ME particle (when such a particle is created) will not be the classical fields, but the quantum (quantum-like) fields. It means that the motion equations inside a local ME particle should be the quantum (quantum-like) motion equations with special symmetry properties.

The fundamental discrete symmetries of parity (P), time reversal (T) and charge conjugation (C), and their violations in certain situations, are central in modern elementary particle physics, and in atomic and molecular physics. As a basic principle, the weak interaction is considered as the only fundamental interaction, which does not respect left-right symmetry. The mutual interaction of magnetic and electric charges in the dynamical construction of the elementary particles could lead in a natural way to the parity violation observed in weak interactions [6]. Atoms are chiral due to the parity-violating weak neutral current interaction between the nucleus and the electrons [7]. Following ideas of some recent theories, one sees that ME interactions in crystal structures with symmetry breakdown arise from special toroidal distributions of currents and are described by so-called anapole moments [8]. The anapole moment takes place in systems with the parity violation and with the annual magnetic field [9]. At present, the role of anapole moments is considered as an important factor in understanding chirality (helicity) in different atomic, molecular and condense-matter phenomena. The anapole moment plays the essential role in nuclear helimagnetism [10, 11]. It was considered as an intrinsic property of a diatomic polar molecule [12].

One of the reasons why the anapole moments appear is Stone's spinning-solenoid Hamiltonian [13] and ME properties of Stone's Hamiltonian become apparent because of Berry's curvature of the electronic wavefunctions. In recent theories of spin waves in magnetic-order crystals, a Berry curvature is stated as playing a key role [14, 15]. The Berry phase may also influence the properties of magnons. If the magnetic medium in which the magnon is propagating is spatially non uniform, a Berry phase may be accumulated along a closed circuit in space. It has been recently indicated [16] that the geometric Berry phase due to a non-coplanar texture of the magnetization of a ferromagnetic ring would affect the dispersion of magnons, lifting the degeneracy of clockwise and anticlockwise propagating magnons. It was found [16] that the magnetization transport by magnons in a noncollinear spin structure is accompanied by an electric polarization. This electric polarization can be experimentally observed not only in the vicinity of the mesoscopic ring [16] but also in the vicinity of the magnetic wire [17]. Moving magnetic dipoles represent an electric dipole moment [18] and are therefore affected by electric fields. Such a ME effect in magnetic nanostructures is, in fact, the relativistic effect of a transformation of magnetization to the electric field in the moving frame. At the same time, because of the electronic structure of the material, the magnitude of such ME coupling can be much larger than a bare relativistic effect [19].

This small survey is not a formal enumeration of basic concepts. The above microscopic and mesoscopic aspects of chirality and magnetoelectricity should, certainly, be related to the problems of local ME particles and the unified ME fields originated from such point sources, and necessarily should become the main subject for realization of local ME composites. One can formulate the concept of particulate ME composites as possible unification of the processes of dipole motions and symmetry breaking phenomena. Following the results of recent studies, we may come to a certain deduction that spectral properties of magnetostatic modes (MSMs), or magnetic-dipolar modes (MDMs) in ferrite disks may put us into a proper way. It was shown that MDMs in a normally magnetized ferrite disk are characterized by the dynamical symmetry breaking effects resulting in ME properties [20 – 25]. Artificial ME materials should be realized as patterned structures composed with special-symmetry ferromagnetic elements.

The purpose of this paper is to analyze interactions between MDM ferrite disks for possible realization of ME "molecules" and local ME composites. We involve a rigorous spectral treatment which uses the two-"atom" localized orbital picture as its basis and show a quantum-like behavior of the ME-particle interactions.

## II. THEORIES OF INTERACTING OSCILLATING ELEMENTS AND A MODEL FOR COUPLED ME PARTICLES

The microwave ME effect in a ferrite disk particle appears due to the vortex states of eigen magnetic-dipolar-mode (MDM) oscillations [23]. The fact that for MDMs in a ferrite disk one has evident both classical and quantum-like attributes [20 – 26], puts special demands on the methods used for study of interacting ferrite ME particles. To develop a proper model for coupled ferrite ME particles we have to make a preliminary analysis of the main aspects related to the subject. This concerns the known models of the magnetization dynamics and the problems of interacting ferromagnetic dots; the methods of the coupled-mode theory for classical and quantum guiding and oscillating systems. As special questions, we have to dwell on the spectral properties of MDM oscillations in a ferrite disk particle and the dynamical symmetry breaking effects of MDM oscillations.

### A. Models for the magnetization dynamics in interacting ferromagnetic elements

At present, studies of resonant modes of structures of interacting ferromagnetic elements (slabs, wires and dots) are a subject of an interest for many researches. Different classical approaches have been developed for such systems. The interparticle coupling, mainly of dipolar nature, affects both the static and dynamic behaviors of the magnetization. For any distribution of magnetization, both for continuum ferromagnetic media and for patterned structures with ferromagnetic elements, the magnetic dipole interaction is described by the magnetostatic solution. If we consider the motion of the magnetization in a particular ferromagnetic element, then a dynamic magnetic dipole field is generated in the spatial region outside the element by the precession of the magnetization. When elements are arranged in the form of a periodic structure, one may seek solutions of the Bloch form.

The effective magnetic dipolar interaction between single domain two-dimensional ferromagnetic particles (magnetic dots) was analyzed in paper [27]. Each particle behaves as a single spin $\vec{S} = N \vec{s}$, where $N$ is the total number of local spins $\vec{s}$ in the particle. The effective interaction between particles of spins $\vec{S}_i$ and $\vec{S}_j$ in a lattice is described as the classical magnetostatic interaction between two magnetic dipoles. As it is discussed in [27], dipolar coupling between particles may induce *ferromagnetic* long range order. Based on a dynamical matrix method, a theory for the determination of the collective spin-wave modes of regular arrays of magnetic particles (taking into account the dipolar interaction between particles) has been developed in [28]. A method of an analysis of magnetic-particle arrays based on an assumption that a body is represented by an array of macrospins, each consisting of many true spins, was developed in [29]. This micromagnetic-simulation method (viewed by the authors of Ref. [29] as a discrete version of the Landau-Lifshitz equation) involves a solution of the coupled Larmor equations of the individual dipoles with all fields acting on them.

Aiming to realization of microwave ME composites, the most interesting aspect for our studies concerns an analysis of the known publications of coupled disks with the vortex states. Magnetically soft ferromagnetic materials generally form domain structures to reduce their magnetostatic (MS) energy. In this context, closure domains are especially suitable. Such magnetic objects are characterized by a closed flux circuit having no magnetic flux leakage outside the material. In very small systems, however, the formation of domain walls is not energetically favored. Specifically, in a dot of a ferromagnetic material of micrometer or submicrometer size, a curling spin configuration – that is, a magnetization vortex – has been proposed to occur in place of domains. The vortex consists of an in-plane, flux-closure magnetization distribution and a central core whose magnetization is perpendicular to the dot plane. It has been shown that under certain conditions a vortex structure will be stable because of competition between the exchange and dipole interactions. For magnetic vortices, one obtains the clockwise (CW) and counter-clockwise (CCW) rotations of magnetization vector $\vec{m}$ in the dot plane [30 – 32].

Two closely spaced vortex-dynamics ferromagnetic disks can be coupled due to the MS interactions. It was shown [33, 34] that the vortex core exhibits circular motion around the disk center. When the vortex core is shifted from a disk center, magnetic charges emerge on the side surface of the disk. Due to these charges one may have the MS interaction between the vortex-dynamics disks [35, 36]. Micromagnetic simulation shows that the coupled vortices coherently rotate around the disk centers and the CW or CCW rotational directions do not influence the dynamics of vortices [35]. Rotational directions of the magnetization play, however, an important role in the vortex coupling in asymmetrical disks. In an isolated perfect circular disk, CW and CCW states are energetically degenerate. By introducing asymmetry in the disk, vortex motion becomes chirality-controlled. In a pair of asymmetrical ferrite disks one has chirality-controlled magnetostatic interactions [37]. The knowledge of the vortex mode structure in an isolated dot allows studies of collective waves for an array of magnetic dots in the vortex state [38]. In paper [38], solutions were obtained with an assumption that dots, having the same states of vorticity and polarization, are coupled via the MS dipolar mechanism of interaction.

**B. Overlap integrals and the coupled mode theories**

In all the papers, to the best of our knowledge, analyzing magnetization dynamics of interacting nanoscale magnetic elements, the problem of interdot coupling is described by the Poisson equation for the magnetostatic potential. At the same time, in numerous spectral problems of interacting eigen oscillating elements, an interaction is considered via evanescent exponential tails of eigen wave functions localized inside a separate element and is described by the overlap integral.

Generally, a form of the overlap integral is determined by the orthogonality conditions of eigen modes in a separate element. This concerns both classical electromagnetic structures and quantum systems. In a case of electromagnetic waveguides, the coupling between adjacent guides induces the transverse dynamics. Energy exchange is caused by the overlap of the evanescent tails of the guided modes. In several coupled-mode formulations for coupled waveguide systems, one has different expressions for the overlap integrals and so different formulas for coupling coefficients [39 – 45].

Methods of the coupled-mode theory are applicable also for an analysis of coupled electromagnetic resonators [46]. Dielectric resonators with extremely high values of the quality factor $Q$ use high-order azimuth oscillations, the so-called whispering-gallery modes [47]. Such modes are, in fact, the Mie-resonance modes of small particles [48]. There are, however, not the modes which arise from the real-norm orthogonality relation of the spectral problem. The peak positions of these resonant modes are dependent on a character of excitation [49]. In spite of the evidently high $Q$ factor of resonators with whispering-gallery modes, all these modes must necessarily be leaky [50]. The analyses of coupled Mie-resonance dielectric particles are made based on phenomenologically introduced overlap integrals, as the analogy with overlap between the modes of coupled quantum wells (atoms). However, as it was noted in [49], the complete analogy between coupled Mie-resonance dielectric particles and coupled quantum particles does not hold because the Mie-resonance states are not enough bound within the dielectric particle. This extended behavior does not guarantee the convergence of the overlap integral between the resonance states of the neighboring particles [49]. Nevertheless, in the theory of coupled whispering-gallery resonators, the overlap is used as a "direct" modal coupling term. The coherent coupling results in the frequency splitting of the corresponding whispering-gallery modes and is a manifestation of the well known phenomena of the normal mode splitting in coupled harmonic oscillators [51].

In quantum systems, an overlap integral is usually defined as the integral over space of the product of the wave function of a particle and the complex conjugate of the wave function of another particle. An analysis of a spectrum of two horizontally coupled 2D quantum dots with two confined electrons is based on the theory of a double-quantum-dot hydrogen molecule. Due to the Coulomb interaction and the Pauli exclusion principle one obtains a highly entangled spin state of two coupled electron wave functions. The exchange coupling between two confined electron states arises as a result of their spatial behavior and can be expressed as an effective spin-spin interaction [52 – 54]. Together with an analysis of the Hilbert space structure of horizontally coupled double-quantum-dot system, a vertically coupled double-quantum-dot system has also been studied [55]. Recently, the concepts of the classical coupled-mode theory were used for coupled electron-wave quantum waveguides [56]. The electron wave propagating in a coupled-quantum-well system is expressed as a linear combination of two guided electron-wave modes in separate quantum waveguides. The overlap integral is determined by the orthogonality conditions of eigen modes in an individual quantum well. An analysis establishes the basic relations between the normal modes of the coupled well system and the isolated modes of the individual wells. It allows calculating the coupling and propagation constants from basic physical quantities of the uncoupled modes.

**C. Energy eigenstates of MDM oscillations and a model for coupled MDM ferrite disks**

In an analysis of the MDM oscillating spectra, a ferrite-disk particle is considered as a section of an axially magnetized ferrite rod. For a flat ferrite disk, having a diameter much bigger than a disk thickness, one can successfully use separation of variables for the MS-potential wave function [26, 57]. A similar way of separation variables is used in solving the electromagnetic-wave spectral problem in dielectric disks [58].

For MDMs in a ferrite disk one has evident quantum-like attributes. The spectrum is characterized by energy eigenstate oscillations which are characterized by a two-dimensional ("in-plane") differential operator

$$\hat{F}_\perp = \frac{g_q}{16\pi} \mu \nabla_\perp^2, \tag{1}$$

where $\nabla_\perp^2$ is the two-dimensional (with respect to cross-sectional coordinates) Laplace operator, $\mu$ is a diagonal component of the permeability tensor, and $g_q$ is a dimensional normalization coefficient for mode $q$. Operator $\hat{F}_\perp$ is positive definite for negative quantities $\mu$. The normalized average (on the RF period) density of accumulated magnetic energy of mode $q$ is determined as

$$E_q = \frac{g_q}{16\pi}\left(\beta_{z_q}\right)^2, \qquad (2)$$

where $\beta_{z_q}$ is the propagation constant of mode $q$ along disk axis $z$. The energy eigenvalue problem is defined by the differential equation:

$$\widehat{F}_\perp \tilde{\eta}_q = E_q \tilde{\eta}_q, \qquad (3)$$

where $\tilde{\eta}_q$ is a dimensionless membrane MS-potential wave function [21, 23]. At a constant frequency, the energy orthonormality for MDMs in a ferrite disk is written as:

$$(E_q - E_{q'})\int_S \tilde{\eta}_q \tilde{\eta}_{q'}^* dS = 0, \qquad (4)$$

where $S$ is a cylindrical cross section of an open disk. One has different mode energies at different quantities of a bias magnetic field. From the principle of superposition of states, it follows that wave functions $\tilde{\eta}_q$ ($q = 1,2,...$), describing our "quantum" system, are "vectors" in an abstract space of an infinite number of dimensions – the Hilbert space. In quantum mechanics, this is the case of so-called energetic representation, when the system energy runs through a discrete sequence of values. In the energetic representation, a square of a modulus of the wave function defines probability to find a system with a certain energy value [59, 60]. In our case, scalar-wave membrane function $\tilde{\eta}$ can be represented as

$$\tilde{\eta} = \sum_q a_q \tilde{\eta}_q \qquad (5)$$

and the probability to find a system in a certain state $q$ is defined as

$$|a_q|^2 = \left|\int_S \tilde{\eta}\, \tilde{\eta}_q^* dS\right|^2. \qquad (6)$$

It was shown [21, 23] that because of the boundary condition on a lateral surface of a ferrite disk, the topological effects are manifested through the generation of relative phases which accumulate on the boundary wave function $\delta_\pm$. There exist the vortex-state resonances which conventionally designated as the (+) and the (–) resonances. For the (+) resonance, a direction of an edge chiral rotation coincides with the precession magnetization direction, while for the (–) resonance, a direction of an edge chiral rotation is opposite to the precession magnetization direction. For a given cross-sectional state (described by the mode membrane function), one defines the strength of a vortex of a whole disk, $s_\pm^e$, and a moment

$$a_\pm^e = i\mu_a\, s_\pm^e, \qquad (7)$$

where $\mu_a$ is an off-diagonal component of the permeability tensor. The superscript "$e$" means "electric" since moment $\vec{a}^e$ has the symmetry of an electric dipole. [21, 23].

MDMs in a normally magnetized ferrite disk are characterized by the dynamical symmetry breakings resulting in the ME effects. A moment $a_\pm^e$ has the anapole moment properties [9, 20 – 24]. From an analysis of the spectral problem for MS-potential wave function it becomes evident that in magnetically saturated cylindrical dots there is a property associated with the vortex structures. The vortices are guaranteed by the chiral edge states of magnetic-dipolar modes in a quasi-2D ferrite disk. Physical nature of such vortices is different from the vortices found in magnetically soft "small" (with the dipolar and exchange energy competition) cylindrical dots [30 – 32].

Spectral properties of MDM oscillations in a ferrite disk determine the basis for elaboration and an analysis of a model for coupled MDM disks. Following Eq. (1), one can see that the energy splitting in coupled MDM disks will be defined by the wavenumber deviations at a constant frequency. This certainly differs from the internal energy splitting in coupled dielectric resonators which is defined by the frequency deviations [51]. Because of eigen electric moments oriented normally to the disk plane, coupling between two ferrite disk particles should be described by the "exchange interaction" overlapping integrals for eigen MS-wave functions. To a certain extent, this can be considered as a dual case with respect to coupled quantum dots with the exchange interaction described by overlapping integrals for eigen electron wave functions [52 – 54]. At the same time, symmetry breaking effects for MDM oscillations result in appearance of special ME interactions for coupled ferrite disks.

Since, in an analysis of the MDM oscillating spectra a disk is represented as a section of an axially magnetized ferrite rod, coupled quasi-2D ferrite disk particles should be analyzed as a section of coupled MDM waveguides. The coupled-mode formulation for MDM ferrite waveguides demands a special consideration. Development of the coupled-mode model for such waveguide structures we should start with consideration of the power flow density for MDMs.

**III. MDM FLOW DENSITY IN A FERRITE ROD**

Any kind of a wave process is characterized by a certain flow density $\vec{j}$. The physical meaning of flow $\vec{j}$ is determined by a type of a differential equation describing a wave process. For Maxwell equations, there is the Poynting vector, for Schrödinger equation, there is the probability flow density. These flows have different physics: in the Maxwell theory, we can define a positive-definite energy density, while cannot define a positive-definite probability density.

In a waveguide structure, there is a longitudinal flow density $\vec{j}_\parallel$, where subscript $\parallel$ means propagation along a waveguide axis. Integration over a waveguide cross section, $J_\parallel \equiv \int_S \vec{j}_\parallel \cdot \vec{e}_z dS$, gives a total flow along a waveguide ($\vec{e}_z$ is the unit vector along $z$ axis). In a case of an electromagnetic waveguide one has the longitudinal electromagnetic power flow density (the Poynting vector) [1]

$$\left(\vec{j}_\parallel\right)_{EM} = \frac{c}{4\pi}\left(\vec{E}_\perp \times \vec{H}_\perp^* + \vec{E}_\perp^* \times \vec{H}_\perp\right), \tag{8}$$

where subscript $\perp$ means transversal field components. For a quantum waveguide, the longitudinal probability flow density is expressed as [59, 60]

$$\left(\vec{j}_\parallel\right)_{QM} = \frac{i\hbar}{2m}\left(\phi \nabla_\parallel \phi^* - \phi^* \nabla_\parallel \phi\right), \tag{9}$$

where $m$ is the electron mass, $\phi$ is the electron wave function, and operator $\nabla_\parallel$ denotes differentiation along a waveguide axis. For a MS-wave waveguide, the longitudinal power flow density is expressed as [23]

$$\left(\vec{j}_\parallel\right)_{MSW} = \frac{i\omega}{16\pi}\left(\psi^* \vec{B}_\parallel - \psi \vec{B}_\parallel^*\right), \qquad (10)$$

where $\psi$ is the MS-potential wave function and $\vec{B}$ is the magnetic flux density. Eq. (10) represents a classical flow density but, at the same time, for a certain configuration, it looks like the probability flow density in a quantum waveguide. Really, for an axially magnetized ferrite rod, one has [23]

$$\left(\vec{j}_\parallel\right)_{MSW} = \frac{i\omega}{16\pi}\left(\psi \nabla_\parallel \psi^* - \psi^* \nabla_\parallel \psi\right). \qquad (11)$$

Let us consider a quantity $\nabla_\parallel \cdot \vec{j}_\parallel$. For a lossless and sourceless linear waveguide one has

$$\nabla_\parallel \cdot \vec{j}_\parallel = 0. \qquad (12)$$

It means conservation of the flow density $j_\parallel$ along a waveguide. For a MS-wave waveguide, divergence $\nabla_\parallel \cdot \vec{j}_\parallel$ is expressed as

$$\nabla_\parallel \cdot \vec{j}_\parallel = \frac{i\omega}{16\pi}\nabla_\parallel \cdot \left(\psi \nabla_\parallel \psi^* - \psi^* \nabla_\parallel \psi\right) = \frac{i\omega}{16\pi}\left(\psi \nabla_\parallel^2 \psi^* - \psi^* \nabla_\parallel^2 \psi\right). \qquad (13)$$

For an open ferrite rod (with homogeneous material parameters), one has the following second-order equations for MS-potential wave function $\psi$. There is the Walker equation inside a ferrite

$$\mu \nabla_\perp^2 \psi + \nabla_\parallel^2 \psi = 0 \qquad (14)$$

and the Laplace equation outside a ferrite

$$\nabla_\perp^2 \psi + \nabla_\parallel^2 \psi = 0. \qquad (15)$$

Based on Eqs. (14) and (15), one rewrites Eq. (13) as

$$\nabla_\parallel \cdot \vec{j}_\parallel = -i\frac{\mu\omega}{16\pi}\left(\psi \nabla_\perp^2 \psi^* - \psi^* \nabla_\perp^2 \psi\right) = -i\frac{\mu\omega}{16\pi}\nabla_\perp \cdot \left(\psi \nabla_\perp \psi^* - \psi^* \nabla_\perp \psi\right) \qquad (16)$$

for an internal ferrite region and

$$\nabla_\parallel \cdot \vec{j}_\parallel = -i\frac{\omega}{16\pi}\left(\psi \nabla_\perp^2 \psi^* - \psi^* \nabla_\perp^2 \psi\right) = -i\frac{\omega}{16\pi}\nabla_\perp \cdot \left(\psi \nabla_\perp \psi^* - \psi^* \nabla_\perp \psi\right) \qquad (17)$$

for an external dielectric region.

Let a ferrite core be a cylinder of radius $\Re$. A necessary requirement of conservation of the power flow density in a lossless regular MS-wave waveguide

$$\nabla_{\|} \cdot \vec{J}_{\|} \equiv \int_S \nabla_{\|} \cdot \vec{j}_{\|} \, dS = 0, \qquad (18)$$

occuring for boundary conditions on a lateral surface of a ferrite rod [21, 23]:

$$\psi_{r=\Re_-} = \psi_{r=\Re_+} \qquad (19)$$

and

$$\mu \left( \frac{\partial \psi}{\partial r} \right)_{r=\Re_-} - \left( \frac{\partial \psi}{\partial r} \right)_{r=\Re_+} = 0 \qquad (20)$$

is in an evident contradiction with another physical requirement, namely, conservation of the magnetic flux density, $\nabla \cdot \vec{B} = 0$. The continuity of a normal component of $\vec{B}$ on a cylindrical surface of a ferrite region takes place if

$$\mu \left( \frac{\partial \psi}{\partial r} \right)_{r=\Re_-} - \left( \frac{\partial \psi}{\partial r} \right)_{r=\Re_+} = i\mu_a \left( \frac{\partial \psi}{\partial \theta} \right)_{r=\Re}. \qquad (21)$$

So one becomes faced with a paradox physical situation that a wave process in a lossless MS-wave waveguide should be accompanied with the "edge anomaly" on a cylindrical surface of a ferrite rod caused by the non-zero term $i\mu_a \left( \frac{\partial \psi}{\partial \theta} \right)_{r=\Re}$. In order to cancel this "edge anomaly", the boundary excitation must be described by chiral states [21, 23]. These chiral states represent an additional degree of freedom resulting in elimination of the "edge anomaly". Referring to the boundary conditions used in variational methods, Eq. (20) corresponds to so called essential boundary conditions, while Eq. (21) corresponds to so called natural boundary conditions [57, 61]. The oscillating modes have the energy orthogonality properties and (due to the edge chiral states) pseudoelectric gauge fields. A flat ferrite disk is considered as a thin section of a ferrite MSW waveguide with an eigen electric moment. This electric moment is described with the spinning (double-valued) coordinates [21, 23].

Now let us consider two parallel identical cylindrical MSW waveguides, *a* and *b*. The ferrite rods are axially magnetized along *z* axis. When we analyze this structure as an *entire* guiding system, equation

$$\nabla_{\|} \cdot \vec{J}_{\|} \equiv \int_{S_\Sigma} \nabla_{\|} \cdot \vec{j}_{\|} \, dS = 0 \qquad (22)$$

(where $S_\Sigma$ is a cross section of a whole two-rod open system) will be satisfied if together with continuity of MS-potential wavefunction on lateral of ferrite rods one has the following boundary conditions for derivatives:

$$\mu\left(\frac{\partial \psi}{\partial r}\right)_{r=\mathfrak{R}_-^a} - \left(\frac{\partial \psi}{\partial r}\right)_{r=\mathfrak{R}_+^a} = 0 \tag{23}$$

and

$$\mu\left(\frac{\partial \psi}{\partial r}\right)_{r=\mathfrak{R}_-^b} - \left(\frac{\partial \psi}{\partial r}\right)_{r=\mathfrak{R}_+^b} = 0. \tag{24}$$

At the same time, the conditions of continuity of a normal component of the magnetic flux density on a cylindrical surface of every ferrite rod are satisfied by two equations similar to Eq. (21). To cancel the "edge anomaly", the boundary excitation must be described by chiral states on cylindrical surfaces of each ferrite rods, *a* and *b*.

An entire structure of two horizontally coupled ferrite disks is considered as a thin section of a two-rod open system of ferrite MSW waveguides with two eigen electric moments. It can be supposed that there should be two separate states: (a) eigen electric moments of interacting ferrite disks are parallel and (b) eigen electric moments of ferrite disks are anti parallel.

## IV. COUPLED-MODE ANALYSIS FOR MDM FERRITE WAVEGUIDES

In a coupled-mode theory, two parallel waveguides are not considered as an *entire* guiding system. This theory is based on an analysis of the individual-waveguide mode interactions and transformations which appear because of the presence of another waveguide. When we put another waveguide parallel and in close vicinity to the first one, the coupling between adjacent guides induces the transverse dynamics. The overlap integrals are the main ingredients in the modal description of the waveguide coupling. These overlap integrals "tell" about the compatibility of interacting modes in both waveguides.

Let us start, however, with a situation when waveguides are placed at *infinite* distance one from another. For every propagating mode in separate waveguide $l$ ( $l = a,b$ ) we have [26]

$$\hat{L}^l V^l = 0, \tag{25}$$

where $\hat{L}^l \equiv \begin{pmatrix} \left(\vec{\vec{\mu}}^l\right)^{-1} & \nabla \\ -\nabla \cdot & 0 \end{pmatrix}$ is the differential-matrix operator, $\vec{\vec{\mu}}$ is the permeability tensor, and $V^l \equiv \begin{pmatrix} \vec{B}^l \\ \psi^l \end{pmatrix}$ is the vector function included in the domain of definition of operator $\hat{L}^l$. Outside of ferrite regions one has the same equations but with $\vec{\vec{\mu}} = \vec{\vec{I}}$, where $\vec{\vec{I}}$ is the unit matrix. Eq. (25) can be rewritten as

$$(\hat{L}_\perp^l - i\beta^l \hat{R})\tilde{V}^l = 0, \tag{26}$$

where $\hat{L}_\perp^l \equiv \begin{pmatrix} \left(\vec{\vec{\mu}}^l\right)^{-1} & \nabla_\perp \\ -\nabla_\perp \cdot & 0 \end{pmatrix}$, subscript $\perp$ means differentiation over a waveguide cross section, $\beta^l$ is the MS-wave propagation constant along $z$ axis, $\tilde{V}^l \equiv \begin{pmatrix} \vec{\tilde{B}}^l \\ \tilde{\psi}^l \end{pmatrix}$ is the membrane vector function ($V^l \equiv \tilde{V}^l e^{-i\beta^l z}$), $\hat{R} \equiv \begin{pmatrix} 0 & \vec{e}_z \\ -\vec{e}_z & 0 \end{pmatrix}$, $\vec{e}_z$ is a unit vector along the axis of the wave propagation.

For *finite* distances between two parallel waveguides, the wave process of every mode in a separate waveguide becomes perturbed by another waveguide. The coupling can be exhibited via perturbation of the power flow $\vec{J}^l_\parallel$ of a separate waveguide *l*. Formally, in this case we can write

$$\hat{L}^l V^l = Q^l, \tag{27}$$

where $Q^l$ are the "source vectors" which will be defined below. In presence of "sources", Eq. (26) should be rewritten as

$$(\hat{L}^l_\perp + \frac{\partial}{\partial z}\hat{R}) V = Q^l. \tag{28}$$

To solve the excitation problem we can use either *complete orthonormal basis* of modes of the guide *a*, or *complete orthonormal basis* of modes of the guide *b*. If the functional basis of waveguide *a* is used, we have

$$V = V^a = \sum_{p=1}^{\infty} a_p(z) \tilde{V}_p^a, \tag{29}$$

where $\tilde{V}_p^a$ is a membrane function of mode *p* in a waveguide *a*. When we use the basis of waveguide *b*, we can write

$$V = V^b = \sum_{q=1}^{\infty} b_q(z) \tilde{V}_q^b, \tag{30}$$

where $\tilde{V}_q^b$ is a membrane function of mode *q* in a waveguide *b*. Based on representation (29) and taking into account Eq. (26), we can write Eq. (28) for waveguide *a* as

$$\sum_{p=1}^{\infty} \left( \frac{1}{a_p(z)} \frac{da_p(z)}{dz} + i\beta_p^a \right) \hat{R} V^a = Q^a, \tag{31}$$

while for waveguide *b* we have

$$\sum_{p=1}^{\infty} \left( \frac{1}{b_q(z)} \frac{db_q(z)}{dz} + i\beta_q^b \right) \hat{R} V^b = Q^b. \tag{32}$$

Here $\beta_p^a$ and $\beta_q^b$ are propagation constants for modes *p* and *q* in unperturbed waveguides *a* and *b*, respectively.

The excitation equation for mode *p* in waveguide *a* is

$$\frac{da_p(z)}{dz} + i\beta_p^a a_p(z) = \frac{1}{N_p^a} \oint_{C^b} Q^a \cdot \left(\widetilde{V}_p^a\right)^* dC \qquad (33)$$

and the excitation equation for mode $q$ in waveguide $b$ is written as

$$\frac{db_q(z)}{dz} + i\beta_q^b b_q(z) = \frac{1}{N_q^b} \oint_{C^a} Q^b \cdot \left(\widetilde{V}_q^b\right)^* dC . \qquad (34)$$

Here $N_p^a \equiv \int_{S^a} \left(\hat{R}\widetilde{V}_p^a\right)\left(\widetilde{V}_p^a\right)^* dS$ and $N_q^b \equiv \int_{S^b} \left(\hat{R}\widetilde{V}_q^b\right)\left(\widetilde{V}_q^b\right)^* dS$ are the norms of modes $p$ and $q$ in waveguides $a$ and $b$, respectively.

What are the "source vectors" $Q^l$? The overlap of the evanescent tails of the guided modes determines the transverse dynamics of the energy exchange in a coupled-waveguide system. When one puts one waveguide in the vicinity of another waveguide, there should be induced sources which make $\nabla_\parallel \cdot \vec{J}_\parallel^l \equiv \int_S \nabla_\parallel \cdot \vec{j}_\parallel^l \, dS \neq 0$ in every separate waveguide. Since no bulk magnetic charges exist, there cannot be any induced "bulk sources" for MS-potential wave functions and their space derivatives. At the same time, we can see that for a separate MS-wave waveguide continuity of the power flow takes place when the boundary condition (20) is satisfied. So it becomes evident that there are induced "surface magnetic sources" caused by fractures of derivatives of MS-potential wave functions of modes $\nu$ of a waveguide $a$ on a surface of waveguide $b$:

$$\left(\rho_s^m\right)^a = \sum_{\nu=1}^{\infty} \left[(1-\mu)\frac{\partial \widetilde{\psi}_\nu^a}{\partial n^b}\right]_{\text{on border } C^b} . \qquad (35)$$

Similarly, the induced "surface magnetic sources" caused by fractures of derivatives of MS-potential wave functions of modes $\chi$ of waveguide $b$ on a surface of waveguide $a$ are expressed as

$$\left(\rho_s^m\right)^b = \sum_{\chi=1}^{\infty} \left[(1-\mu)\frac{\partial \widetilde{\psi}_\chi^b}{\partial n^a}\right]_{\text{on border } C^a} . \qquad (36)$$

In Eqs. (35) and (36), $n^a$ and $n^b$ are external normals to border contours $C^a$ and $C^b$, respectively. We suppose that ferrite rods $a$ and $b$ are characterized by the same material parameters. It is necessary to note also that in an axially magnetized ferrite rod, MDMs propagate at negative quantity $\mu$ [26, 57].

For coupled MDM waveguides the "source vector" $Q^l$ is expressed as

$$Q^l \equiv i \begin{pmatrix} 0 \\ \left(\rho_s^m\right)^l \end{pmatrix}, \qquad (37)$$

As a result, we rewrite Eqs. (33) and (34) as

$$\frac{da_p(z)}{dz} + i\beta_p^a a_p(z) = i\frac{1}{N_p^a} \oint_{C^b} \left\{ \sum_{\nu=1}^{\infty} \left[ (1-\mu)\frac{\partial \tilde{\psi}_\nu^a}{\partial r} \right]_{\text{on border } C^b} \right\} \left(\tilde{\psi}_p^a\right)^* dC \qquad (38)$$

and

$$\frac{db_q(z)}{dz} + i\beta_q^b b_q(z) = i\frac{1}{N_q^b} \oint_{C^a} \left\{ \sum_{\chi=1}^{\infty} \left[ (1-\mu)\frac{\partial \tilde{\psi}_\chi^b}{\partial r} \right]_{\text{on border } C^a} \right\} \left(\tilde{\psi}_q^b\right)^* dC. \qquad (39)$$

In the theory of coupled waveguide structures, one usually restricts an analysis with consideration of two modes in separate waveguides. Following this idea of a coupled-mode model we express the total field $V$ as a linear combination of two guided modes in waveguides $a$ and $b$:

$$V \approx A_p(z)\tilde{V}_p^a + B_q(z)\tilde{V}_q^b. \qquad (40)$$

Based on this approximate representation, we write the excitation equations for modes $p$ and $q$ as

$$\frac{da_p(z)}{dz} + i\beta_p^a a_p(z) = iK_{pp}^{aa} A_p(z) + iK_{pq}^{ab} B_q(z), \qquad (41)$$

$$\frac{db_q(z)}{dz} + i\beta_q^b b_q(z) = iK_{qq}^{bb} B_q(z) + iK_{qp}^{ba} A_q(z), \qquad (42)$$

where

$$K_{pp}^{aa} = \frac{1}{N_p^a} \oint_{C^b} \left[ (1-\mu)\frac{\partial \tilde{\psi}_p^a}{\partial r} \right] \left(\tilde{\psi}_p^a\right)^* dC, \qquad (43)$$

$$K_{pq}^{ab} = \frac{1}{N_p^a} \oint_{C^b} \left[ (1-\mu)\frac{\partial \tilde{\psi}_q^b}{\partial r} \right] \left(\tilde{\psi}_p^a\right)^* dC, \qquad (44)$$

$$K_{qq}^{bb} = \frac{1}{N_q^b} \oint_{C^a} \left[ (1-\mu)\frac{\partial \tilde{\psi}_q^b}{\partial r} \right] \left(\tilde{\psi}_q^b\right)^* dC, \qquad (45)$$

$$K_{qp}^{ba} = \frac{1}{N_q^b} \oint_{C^a} \left[ (1-\mu)\frac{\partial \tilde{\psi}_p^a}{\partial r} \right] \left(\tilde{\psi}_q^b\right)^* dC. \qquad (46)$$

On the basis of representations (29), (30) and using the mode orthogonality relations [26], we obtain

$$a_p(z) = A_p(z) + C_{pq}^{ab} B_q(z), \qquad (47)$$

$$b_q(z) = B_q(z) + C_{qp}^{ba} A_p(z), \tag{48}$$

where coefficients $C_{pq}^{ab}$ and $C_{qp}^{ba}$ describe the mode overlap

$$C_{pq}^{ab} = \frac{1}{N_p^a} \int_S \left(\hat{R}\tilde{V}_q^b\right)\left(\tilde{V}_p^a\right)^* dS, \tag{49}$$

$$C_{qp}^{ba} = \frac{1}{N_q^b} \int_S \left(\hat{R}\tilde{V}_p^a\right)\left(\tilde{V}_q^b\right)^* dS, \tag{50}$$

After some manipulations we have the coupled-mode equations

$$\frac{dA_p(z)}{dz} = -i\delta^a A_p(z) + ik^{ab} B_q(z), \tag{51}$$

$$\frac{dB_q(z)}{dz} = -i\delta^b B_q(z) + ik^{ba} A_q(z), \tag{52}$$

where

$$\delta^a = \beta_p^a + \frac{K_{pp}^{aa} - C_{pq}^{ab} K_{qp}^{ba} + C_{pq}^{ab} C_{qp}^{ba}\left(\beta_q^b - \beta_p^a\right)}{1 - C_{pq}^{ab} C_{qp}^{ba}}, \tag{53}$$

$$\delta^b = \beta_q^b + \frac{K_{qq}^{bb} - C_{qp}^{ba} K_{pq}^{ab} + C_{pq}^{ab} C_{qp}^{ba}\left(\beta_p^a - \beta_q^b\right)}{1 - C_{pq}^{ab} C_{qp}^{ba}}, \tag{54}$$

$$k^{ab} = \frac{K_{pq}^{ab} + C_{pq}^{ab}\left(\beta_q^b - \beta_p^a\right) - C_{pq}^{ab} K_{qq}^{bb}}{1 - C_{pq}^{ab} C_{qp}^{ba}}, \tag{55}$$

$$k^{ba} = \frac{K_{qp}^{ba} + C_{qp}^{ba}\left(\beta_p^a - \beta_q^b\right) - C_{qp}^{ba} K_{pp}^{aa}}{1 - C_{pq}^{ab} C_{qp}^{ba}}. \tag{56}$$

Assuming that solutions of Eqs. (51), (52) are proportional to $\exp(-i\vartheta z)$, where $\vartheta$ is a propagation constant of an entire two-rod guiding system, one obtains from the characteristic equation:

$$\vartheta = \frac{\delta^a + \delta^b}{2} \pm \sqrt{\left(\frac{\delta^a - \delta^b}{2}\right)^2 - k^{ab} k^{ba}}. \tag{57}$$

There are propagation constants for eigen modes in an entire guiding system.

For further coupled-mode analysis we will consider only the case when two separate ferrite rods have identical parameters and identical modes, that is, in Eqs. (51) – (57) we use: $\beta_p^a = \beta_q^b$. For a given type of a mode in a separate rod, we have two solutions for propagation constant $\vartheta$ in a coupled-rod system. These solutions correspond to symmetrical and anti-symmetrical field distributions for membrane functions in ferrite rods.

Based on our approach for a single MDM ferrite disk [26, 57], we can analyze literally coupled MDM ferrite disks as a section of coupled MDM waveguides. In such a model (which is applicable for ferrite disks with big diameter-to-thickness ratios), one obtains eigen wavenumbers of oscillating modes as a result of joint solutions of two equations: (a) Eq. (57) for a two-rod guiding system and (b) a transcendental equation for a normally magnetized ferrite film

$$\tan(\vartheta h) = -\frac{2\sqrt{-\mu}}{1+\mu}, \tag{58}$$

where $h$ is thickness of a disk; $\mu$ is a negative quantity. Based on solutions of Eqs. (57), (58) and taking into account Eq. (1), one has the energy for symmetrical *(S)* and anti-symmetrical *(A)* MDM modes in coupled ferrite disks:

$$E^{(S,A)} = \frac{g^{(S,A)}}{16\pi}\left(\vartheta^{(S,A)}\right)^2. \tag{59}$$

## V. IDENTITY AND "EXCHANGE" INTERACTION OF MDM FERRITE DISKS

Two laterally interacting ferrite samples are considered as *identical* particles when a separate MDM disk cannot be clearly distinctive as the "left" or "right" one. We will show that the fact of identity of two MDM ferrite disks depends on a combined effect of symmetry properties of the single-valued membrane wave function, double-valued edge wave function, and a direction of the RF magnetization precession.

Following our previous notations (see e.g. [57]) we represent a membrane function for a certain mode $p$ as

$$\tilde{\psi}_p = C_p \tilde{\varphi}_p, \tag{60}$$

where $C_p$ is a dimensional normalization coefficient and $\tilde{\varphi}_p$ is a dimensionless membrane function. At the same time, for a ferrite disk with $r$ and $\theta$ in-plane coordinates, the MS-potential membrane function $\tilde{\varphi}$ is represented as a product of two functions [21, 23]:

$$\tilde{\varphi} = \tilde{\eta}(r,\theta)\,\delta_\pm, \tag{61}$$

where $\tilde{\eta}(r,\theta)$ is a single-valued membrane function, and $\delta_\pm$ is a double-valued edge (spin-coordinate-like) function. The function $\tilde{\eta}(r,\theta)$ [which satisfies, in fact, the boundary condition (20)] defines the energy eigen states in a ferrite disk, while the topological effects in the MDM ferrite disk are manifested through the generation of relative phases which accumulate on the boundary wave functions $\delta_\pm \equiv f_\pm e^{-iq_\pm \theta}$. For better understanding the topological properties of MDM oscillations, we may introduce also a "spin variable" $\theta'$, defining the orientation of the "spin moment" and two double-valued wave functions, $\delta_+(\theta')$ and $\delta_-(\theta')$, the former corresponding to the eigen value $q_+ = +l\frac{1}{2}$ and the latter to the eigen value $q_- = -l\frac{1}{2}$, where $l = 1, 3, 5, \ldots$ The two wave functions are normalized and mutually orthogonal, so that they satisfy the equations $\int \delta_+^2(\theta') d\theta' = 1$, $\int \delta_-^2(\theta') d\theta' = 1$, and $\int \delta_+(\theta') \delta_-(\theta') d\theta' = 0$. A membrane wave function $\tilde{\varphi}$ is then a function of three coordinates, two positional coordinates such as $r, \theta$, and the "spin coordinate" $\theta'$. For the positional wave function $\tilde{\eta}(r,\theta)$, there could be two *equiprobable* solutions for the membrane wave functions: $\tilde{\varphi}_+ = \tilde{\eta}(r,\theta)\delta_+(\theta')$ and $\tilde{\varphi}_- = \tilde{\eta}(r,\theta)\delta_-(\theta')$.

For a ferrite disk of radius $\Re$, circulation of gradient $\vec{\nabla}_\theta \delta_\pm = -i\frac{q_\pm f_\pm}{\Re} e^{-iq_\pm \theta} \vec{e}_\theta$ along a disk border contour $C = 2\pi\Re$ gives a nonzero quantity when $q_\pm$ is a number divisible by $\frac{1}{2}$. The quantity $\nabla_\theta \delta_\pm$ is defined as the velocity of an irrotational "border" flow: $(\vec{v}_\theta)_\pm \equiv \vec{\nabla}_\theta \delta_\pm$. In such a sense, functions $\delta_\pm$ are the velocity potentials. Circulation of $(\vec{v}_\theta)_\pm$ along a contour $C$ is equal to $\oint_C (\vec{v}_\theta)_\pm \cdot d\vec{C} = \Re \int_0^{2\pi} \nabla_\theta \delta_\pm \, d\theta = \Re \int_0^\pi \nabla_{\theta'} \delta_\pm \, d\theta' = -2f_\pm$. Taking into account that the total MS-potential function $\psi$ is represented as a product: $\psi = \tilde{\psi}\, \xi(z)$ [57], where $\xi(z)$ is the function characterizing $z$-distribution of the MS potential in a ferrite disk, we define the "spin moment" of a whole ferrite disk as

$$\sigma_\pm^e \equiv \int_0^h \xi(z) dz \oint_C (\vec{v}_\theta)_\pm \cdot d\vec{C} = -2f_\pm \int_0^h \xi(z) dz. \tag{62}$$

In a case of a cylindrical ferrite disk, a single-valued membrane function is represented as $\tilde{\eta}(r,\theta) = R(r)\phi(\theta)$, where $R(r)$ is described by the Bessel functions and $\phi(\theta) \sim e^{-i\nu\theta}$, $\nu = \pm 1, \pm 2, \pm 3 \ldots$. Taking into account the "orbital" function $\phi(\theta)$, we may consider the quantity $[\nabla_\theta(\tilde{\eta}\,\delta_\pm)]_{r=\Re}$ as the total ("orbital" and "spin") velocity of an irrotational "border" flow:

$$(\vec{V}_\theta)_\pm \equiv [\nabla_\theta(\tilde{\eta}\,\delta_\pm)]_{r=\Re} = (\tilde{\eta}\nabla_\theta\delta_\pm + \delta_\pm\nabla_\theta\tilde{\eta})_{r=\Re} = -i\frac{(\nu+q_\pm)}{\Re} R_{r=\Re} f_\pm e^{-i(\nu+q_\pm)\theta} \vec{e}_\theta. \tag{63}$$

We define the strength of the total ("orbital" and "spin") vortex of a whole disk as

$$s_\pm^e \equiv R_{r=\Re} \int_0^h \xi(z) dz \oint_C (\vec{V}_\theta)_\pm \cdot d\vec{C} = \Re R_{r=\Re} \int_0^h \xi(z) dz \int_0^{2\pi} (\vec{V}_\theta)_\pm \cdot \vec{e}_\theta \, d\theta = -2 f_\pm R_{r=\Re} \int_0^h \xi(z) dz. \tag{64}$$

This circulation around a lateral border of a ferrite disk is a non-zero quantity because of the presence of double-valued edge (spin-coordinate-like) functions $\delta_\pm$ (it is evident that the circulation is non-zero due to the term $\tilde{\eta}\vec{\nabla}_\theta\delta_\pm$, while the circulation of the term $\delta_\pm\vec{\nabla}_\theta\tilde{\eta}$ is equal to zero). It is important to note that the circulation integral of function $(\vec{V}_\theta)_\pm$ and therefore a quantity of $s_\pm^e$ do not depend on the azimuth phase relation between functions $\tilde{\eta}(\theta)$ and $\delta_\pm(\theta)$.

The quantity $(\vec{V}_\theta)_\pm$ has a clear physical meaning. In the spectral problem for MDM ferrite disks, the border term $-i\mu_a(H_\theta)_{r=\Re}$ arises from the demand of conservation of the magnetic flux density.

Circulation of this border term defines a moment $\vec{a}^e_\pm$ which is expressed by Eq. (7). This moment can be formally represented as a result of a circulation of a quantity, which we call a density of an effective boundary magnetic current $\vec{i}^m$:

$$a^e_\pm = 4\pi \int_0^h \xi(z)\, dz \oint_C \vec{i}^m_\pm \cdot d\vec{C}, \qquad (65)$$

where $\vec{i}^m_\pm \equiv \rho^m (\vec{V}_\theta)_\pm$ and $\rho^m \equiv i \frac{\mu_a}{4\pi} \xi R_{r=\Re}$. In our continuous-medium model, a character of the magnetization motion becomes apparent via the gyration parameter $\mu_a$ in the boundary term for the spectral problem. There is the magnetization motion through a non-simply-connected region. On the edge region, the chiral symmetry of the magnetization precession is broken to form a flux-closure structure. The edge magnetic currents can be observable only via its circulation integrals, not pointwise. This results in the moment oriented along a disk normal. As it was shown experimentally, such a moment has a response in an external RF electric field [24, 62]. The eigen electric moments of a ferrite disk arises not from the classical curl electric fields of magnetostatic oscillations. At the same time, any induced electric polarization effects in a ferrite material are beyond the frames of the experimentally observed multiresonance spectra. An electric moment $a^e_\pm$ is characterized by the anapole-moment properties. This is a certain-type toroidal moment. Some important notes should be given here to characterize properties of moment $\vec{a}^e$. From classical consideration it follows that for a given electric current $\vec{i}^e$, a magnetic dipole moment is described as $\vec{M} = \frac{1}{2c} \int \vec{r} \times \vec{i}^e\, dv$, while the toroidal dipole moment is described as $\vec{t} = \frac{1}{3c} \int \vec{r} \times (\vec{r} \times \vec{i}^e)\, dv$ (see e.g. [63]). When we introduce the notion of an elementary magnet: $\vec{M}_{elem} \equiv \vec{r} \times \vec{i}^e$, we can represent the toroidal dipole moment as a linear integral around a loop: $\vec{t} = \frac{1}{3c} \int \vec{r} \times \vec{M}_{elem}\, dl$. It is considered as a ring of elementary magnets. In this formulation, it is clear that a toroidal moment is parity odd and time reversal odd. In a case when $\vec{m}$ is time varying (due to precession), one has a magnetic current $\vec{i}^m \sim \frac{\partial M_{elem}}{\partial t} \vec{e}_\theta$, where $\vec{e}_\theta$ is a unit vector along a tangent of a loop. A linear integral of this current around a loop defines a moment which is parity odd and time reversal even. For oscillating MDMs one has the azimuth varying border-loop magnetic current (see Fig. 3 in Ref. [23]). The magnetic current $i^m$ is described by the double valued functions. This results in appearance of an anapole moment $\vec{a}^e$, which has the symmetry of an electric dipole – the parity-odd and time-reversal even properties.

Let us choose the azimuth phase relation between functions $\tilde{\eta}(\theta)$ and $\delta_\pm(\theta')$ so that a maximum (minimum) of function $\tilde{\eta}(\theta)$ corresponds to zero of function $\delta_\pm(\theta')$. Following this choice of the azimuth phase relation between functions $\tilde{\eta}(\theta)$ and $\delta_\pm(\theta')$ in a disk, let us consider now two separate identical-parameter ferrite disks with the same direction of a normal bias magnetic field (i.e. with the same direction of the RF magnetization precession $\vec{m}$). Let membrane wave functions $\tilde{\eta}$ of these disks are mutually shifted in phase by $180°$ and the disks are characterized by different double-valued wave functions, $\delta_+$ and $\delta_-$. Since a difference between eigen values $q_+$ and $q_-$ of double-valued wave functions $\delta$ is an integer quantity, it is evident that such disks are absolutely identical. There are, however, two cases of the disk identity. Fig. 1 (a) illustrates the first case of the disk identity. The types of membrane wave functions $\tilde{\eta}$ of these disks are conventionally represented as combinations of two different colour-texture spots on a disk surface. The double-valued wave functions, $\delta_+$ and $\delta_-$, are conventionally shown by arrows. There are also shown orientations of vectors $\vec{\sigma}^e$, $\vec{s}^e$, and $\vec{a}^e$. Below the pictures of ferrite disks one sees the graphs of functions $\delta_\pm(\theta')$, $\vec{\nabla}_{\theta'}\delta_\pm(\theta')$ and $\tilde{\eta}(\theta)$. The second case of the disk identity is shown in Fig. 1 (b). This case has another correlation between signs of functions $\tilde{\eta}$ and $\delta$. Following the pictures of the $\tilde{\eta}$ - and $\delta$ -function distributions, it is worth noting that for two cases shown in Figs. 1 (a) and 1 (b) one has for identical disks coinciding directions of vectors $\vec{s}^e$ and $\vec{a}^e$, and opposite directions of vectors $\vec{\sigma}^e$.

Now let us consider two laterally coupled disks. The disks have identical parameters and are biased by the same DC magnetic field. Following the above coupled-mode theory, one has symmetrical $\tilde{\eta}^{(S)}$ and anti-symmetrical $\tilde{\eta}^{(A)}$ solutions for the positional wave function in the coupled-disk system. The energy splitting is defined by the wavenumber deviation between $\chi^{(S)} \equiv \chi(\tilde{\eta}^{(S)})$ and $\chi^{(A)} \equiv \chi(\tilde{\eta}^{(A)})$ at a constant frequency [see Eq. (59)]. It becomes evident that in a case of a symmetrical $\tilde{\eta}^{(S)}$ solution, two neighboring disks have the azimuth-coordinate MS-potential-distribution pictures shifted to $\pi$. No such a shift one has in a case of an antisymmetrical $\tilde{\eta}^{(A)}$ solution. For a coupled structure, two disks should be identical when one simultaneously exchanges positional coordinates and "spin coordinates". Because of symmetry properties of edge-function chiral rotations, it means that a symmetrical $\tilde{\eta}^{(S)}$ solution will be associated with anti-symmetrical edge-function chiral rotations, and conversely, an anti-symmetrical $\tilde{\eta}^{(A)}$ solution will be associated with symmetrical edge-function chiral rotations. These cases are illustrated in Figs. 2 and 3, respectively. Taking into account the spin coordinates one sees that in a case of Fig. 2 there are opposite directed "spins moments" $\vec{\sigma}^e$ of two disks, while in a case of Fig. 3 the "spins moments" have the same directions. The above situation clearly resembles the Pauli principle for two electrons in the hydrogen molecule: the total (taking into account the positional and spin coordinates) wave function must be antisymmetric with respect to the simultaneous interchange of the coordinates and of the spin variables of the electrons. For two different states of the "spin moment" orientations there are two different "exchange" energies of the "molecule": $E_{\uparrow\uparrow}$ and $E_{\uparrow\downarrow}$, where arrows show directions of vectors $\vec{\sigma}^e$. For the "molecule" with the "exchange" energy $E_{\uparrow\uparrow}$ a total "spin moment" is equal to an odd integer quantity, while for the "molecule" with the energy $E_{\uparrow\downarrow}$ the total "spin order" is zero or an even integer quantity. With increasing distance between disk centers, the overlap between the disk membrane functions falls off exponentially resulting in rapid decrease the "exchange" energy. The "exchange" interaction between MDM ferrite disks is not the same as the magnetostatic interaction between magnetic dipoles.

## VI. ON THE ELECTRIC INTERACTION BETWEEN LATERALLY COUPLED MDM FERRITE DISKS

The "exchange" interaction between identical disks is connected with necessary correlation appearing because of the "spin" symmetrization of the MDM wave functions. Because of existing pseudo-electric fluxes in MDM ferrite disks, an electric interaction has to be taken also into consideration. The MDM electric interaction implies the ability of the edge function in one location to produce phase accumulation in the edge function in another location. The physics of such an interaction is based on the Aharonov-Bohm effect.

In accordance with the spectral analysis in Ref. [23] it follows that the flux of the pseudo-electric field in a MDM ferrite disk arises from necessity to preserve the single-valued nature of the membrane functions. For a separate particle, to compensate for sign ambiguities and thus to make wave functions single valued we added a vector-potential-type term to the MS-potential Hamiltonian. A circulation of vector $\vec{A}_\theta^m$ should enclose a certain flux. The corresponding flux of pseudo-electric field $\vec{\in}$ (the gauge field) through a circle of radius $\Re$ is obtained as:

$$\oint_C (\vec{A}_\theta^m)_\pm \cdot d\vec{C} = \int_S (\vec{\in})_\pm \cdot d\vec{S} = (\Xi^e)_\pm = 2\pi q_\pm, \quad (66)$$

where $(\Xi^e)_\pm$ is the flux of pseudo-electric field. There should be the positive and negative fluxes. These different-sign fluxes should be inequivalent to avoid the cancellation [23, 64]. For non-interacting (placed at infinite distance one from another) identical ferrite disks, *a* and *b*, one has pseudo-electric fluxes $(\Xi^e)_p^a$ and $(\Xi^e)_p^b$ for a given MDM *p*:

$$i\Re^{a,b} \int_0^{2\pi} [(\vec{\nabla}_\theta \delta^{a,b})(\delta^{a,b})^*]_{r=\Re_{a,b}} d\theta = \oint_{C^{a,b}} (\vec{A}_\theta^m)_p^{a,b} \cdot d\vec{C} = \int_{S^{a,b}} (\vec{\in})^{a,b} \cdot d\vec{S} = (\Xi^e)^{a,b} = 2\pi q. \quad (67)$$

Here and further we omit the signs $\pm$.

Now let us take into account a possible electric interaction for two laterally coupled MDM disks. In an assumption about the ability of the edge function in one location to produce phase accumulation in the edge function in another location, the electric interaction presumes an existence of four pseudo-electric fluxes. We will designate a pseudo-electric flux penetrating the border loop of disk *a* as $(\Xi^e)_p^{aa}$ which is connected with a double-valued edge function $\delta^{aa}$ via the Berry connections $(\vec{A}_\theta^m)_p^{aa}$ as

$$i\Re^a \int_0^{2\pi} [(\vec{\nabla}_\theta \delta^{aa})(\delta^{aa})^*]_{r=\Re^a} d\theta = \oint_{C^a} (\vec{A}_\theta^m)_p^{aa} \cdot d\vec{C} = (\Xi^e)_p^{aa} \quad (68)$$

and a pseudo-electric flux penetrating the border loop of disk *b* as $(\Xi^e)_p^{bb}$ which is connected with a double-valued edge function $\delta^{bb}$ via the Berry connections $(\vec{A}_\theta^m)_p^{bb}$ as

$$i\Re^b \int_0^{2\pi} [(\vec{\nabla}_\theta \delta^{bb})(\delta^{bb})^*]_{r=\Re^b} d\theta = \oint_{C_b} (\vec{A}_\theta^m)_p^{bb} \cdot d\vec{C} = (\Xi^e)_p^{bb}. \quad (69)$$

At the same time, we will designate a pseudo-electric flux connected with an edge function of a ferrite disk *b* and penetrating the border loop of disk *a* as $(\Xi^e)_p^{ab}$ and a pseudo-electric flux connected with an edge function of a ferrite disk *a* and penetrating the border loop of disk *b* as $\left(\Xi^e\right)_p^{ba}$:

$$i\Re^a \int_0^{2\pi} [(\vec{\nabla}_\theta \delta^{ab})(\delta^{ab})^*]_{r=\Re^a} d\theta = \oint_{C_a} \left(\vec{A}_\theta^m\right)_p^{ab} \cdot d\vec{C} = \left(\Xi^e\right)_p^{ab} \tag{70}$$

and

$$i\Re^b \int_0^{2\pi} [(\vec{\nabla}_\theta \delta^{ba})(\delta^{ba})^*]_{r=\Re^b} d\theta = \oint_{C_b} \left(\vec{A}_\theta^m\right)_p^{ba} \cdot d\vec{C} = \left(\Xi^e\right)_p^{ba}. \tag{71}$$

Such fluxes are connected with a double-valued edge functions $\delta^{ab}$ and $\delta^{ba}$ via the Berry connections $\left(\vec{A}_\theta^m\right)_p^{ab}$ and $\left(\vec{A}_\theta^m\right)_p^{ba}$, respectively.

The above theory of "exchange" interaction is based on an assumption that an interaction between MDM ferrite disks is enough weak so that a mode portrait of a membrane function in every disk, $\tilde{\eta}_p^a(r,\theta)$ and $\tilde{\eta}_p^b(r,\theta)$, does not change and is the same as in a separate particle. This conservation of mode portraits of membrane functions presumes the conservation of "spin moments" $\left(\vec{\sigma}^e\right)^a$ and $\left(\vec{\sigma}^e\right)^b$ of interacting disks. From this statement it follows that to preserve the singlevaluedness of the membrane function of disks *a* and *b* we have

$$\left(\Xi^e\right)_p^{aa} + \left(\Xi^e\right)_p^{ab} = \left(\Xi^e\right)_p^a = 2\pi q \tag{72}$$

and

$$\left(\Xi^e\right)_p^{bb} + \left(\Xi^e\right)_p^{ba} = \left(\Xi^e\right)_p^b = 2\pi q. \tag{73}$$

It means that fluxes $\left(\Xi^e\right)_p^{aa}$, $\left(\Xi^e\right)_p^{ab}$, $\left(\Xi^e\right)_p^{ba}$, and $\left(\Xi^e\right)_p^{bb}$ are not characterized by discrete quantities. Because the linearity and reciprocity of ME interaction, we can write

$$\left(\Xi^e\right)_p^{ab} = k_p \left(\Xi^e\right)_p^b = 2k_p \pi q \tag{74}$$

and

$$\left(\Xi^e\right)_p^{ba} = k_p \left(\Xi^e\right)_p^a = 2k_p \pi q, \tag{75}$$

where $k_p$ is the ME interaction coefficient for mode *p*. For mode *p*, coefficient *k* determines a fraction of a total pseudo-electric flux of disk *a* perceiving the border ring of disk *b* and, equally, a fraction of a total pseudo-electric flux of disk *b* perceiving the border ring of disk *a*. Evidently, $0 \leq k_p \leq 1$. From the above equations one has evident relations:

$$\left(\Xi^e\right)_p^{aa} = \left(\Xi^e\right)_p^{bb} \tag{76}$$

and

$$\left(\Xi^e\right)_p^{ab} = \left(\Xi^e\right)_p^{ba}. \tag{77}$$

It is useful to note that Eqs. (72) and (73) correspond to the following integral relations:

$$\int_0^{2\pi} [(\vec{\nabla}_\theta \delta^{aa})(\delta^{aa})^*]_{r=\Re_a} d\theta + \int_0^{2\pi} [(\vec{\nabla}_\theta \delta^{ab})(\delta^{ab})^*]_{r=\Re_a} d\theta = \int_0^{2\pi} [(\vec{\nabla}_\theta \delta^a)(\delta^a)^*]_{r=\Re_a} d\theta = 2\pi q \tag{78}$$

and

$$\int_0^{2\pi} [(\vec{\nabla}_\theta \delta^{bb})(\delta^{bb})^*]_{r=\Re_b} d\theta + \int_0^{2\pi} [(\vec{\nabla}_\theta \delta^{ba})(\delta^{ba})^*]_{r=\Re_b} d\theta = \int_0^{2\pi} [(\vec{\nabla}_\theta \delta^b)(\delta^b)^*]_{r=\Re_b} d\theta = 2\pi q. \tag{79}$$

The pseudo-electric-flux interaction resulting in redistributions of the double-valued edge functions $\delta$ has no direct influence on the "exchange"-interaction mechanism of identical MDM ferrite disks. Nevertheless, redistributions of the edge functions $\delta$ will lead to redistributions of edge magnetic currents in ferrite disks. This can be considered as a certain mechanism of interactions between anapole moments $\vec{a}^e$ of the disks. When the flux of the pseudo-electric field in a MDM ferrite disk arises from necessity to preserve the single-valued nature of the membrane functions, the loop magnetic current arises from the demand of conservation of the magnetic flux density on a border surface of a disk. These effects of conservation, being mutually correlated, are important for an analysis of the MDM interactions. Our method, where the MS-potential wave function of a ME "molecule" is created based on the MS-potential wave function of isolated MDM ferrite disks, can be considered as the first-order interaction approach. Taking into account interactions between anapole moments $\vec{a}^e$ of the disks is considered as the second-order interaction approximation. This second-order approximation is beyond the frames of the present paper and should be a subject for a future analysis.

## VII. DISCUSSION AND CONCLUSION

In this paper, we presented theoretical studies of spectral properties of literally coupled of MDM ME disks. We showed that there exists the "exchange" mechanism of interaction between the particles, which is distinctive from the magnetostatic interaction between magnetic dipoles.

In a quasi-2D ferrite disk with a dominating role of magnetic-dipolar spectra, the oscillating spectrum is characterized by energy eigenstates. Because of the strong influence which the boundary geometry has on the energy spectrum of the MDMs, the eigen MS-potential functions are characterized by the vortex states. The vortices are guaranteed by the chiral edge states which result in appearance of eigen electric moments oriented normally to the disk plane.

The "exchange" interaction between coupled MDM ME particles does not represent, certainly, a dual case with respect to the real exchange interaction between coupled natural complex atoms. One of the main distinctive factors concerns the symmetry breaking effects of MDM oscillations. MDM ferrite disks have evident chiral states and the near fields of such particles are characterized by very specific symmetry properties. Nevertheless, a certain resemblance with interacting natural atoms and interacting MDM ferrite disks can be found.

In the simple Heisenberg model each atom is thought to have a single electron which interacts with its neighbor and the dominant interaction is considered to arise from a superposition of this two-electron interaction. Similarly to the Heisenberg mathematical description of the exchange interaction, we can formally introduce the operator characterizing "exchange" interaction between MDM ferrite disks:

$$\hat{\aleph}^e_{"exchange"} = -\Im^e \vec{\sigma}^e_1 \cdot \vec{\sigma}^e_2, \tag{80}$$

where $\Im^e$ is a certain function of $\vec{r}$ – the distance between disk centers – which is chosen so that the eigenvalues of the operator $\hat{\aleph}^e_{"exchange"}$ (in the space of the "spin variables") are equal to the energies $E_{\uparrow\uparrow}$ and $E_{\uparrow\downarrow}$. In Eq. (80), superscripts "$e$" means "electric" [21, 23]. The MDM ferrite disks can be coupled to form "artificial molecules" or an extended superlattice. Such periodic arrays of coupled MDM dots – "artificial crystals" – are interesting both on a fundamental level and from a more application-oriented point of view. When we assume that all the disks in an "artificial crystal" have $q = \frac{1}{2}$, we can write for a lattice:

$$\hat{\aleph}^e_{"exchange"} = -\frac{1}{2}\sum_{l \neq m} \Im^e(\vec{r}_{lm}) \vec{\sigma}^e_l \cdot \vec{\sigma}^e_m. \tag{81}$$

In Section VI of the paper, we pointed out that interaction between anapole moments of the disks can be considered as the second approximation in the analysis. We may treat this anapole-anapole type of interaction similar to the classical dipole-dipole interaction between electric dipoles [1]. Such an electric-dipole-like interaction described by an operator

$$\hat{\aleph}^e_{anapole-anapole} \propto \sum_{l \neq m} \frac{1}{r^5_{lm}} \left[ \left(\vec{a}^e_l \cdot \vec{a}^e_m\right) r^2_{lm} - 3\left(\vec{a}^e_l \cdot \vec{r}_{lm}\right)\left(\vec{a}^e_m \cdot \vec{r}_{lm}\right) \right] \tag{82}$$

should be considered as a characteristic additional to the "exchange" interaction.

**Figure captions**

Fig. 1. Two cases, (a) and (b), of identical MDM ferrite disks with different distributions of membrane functions $\tilde{\eta}$ and edge functions $\delta$.

Fig. 2. Two cases of coupled MDM ferrite disks with $\tilde{\eta}^{(S)}$ solutions.

Fig. 3. Four cases of coupled MDM ferrite disks with $\tilde{\eta}^{(A)}$ solutions.

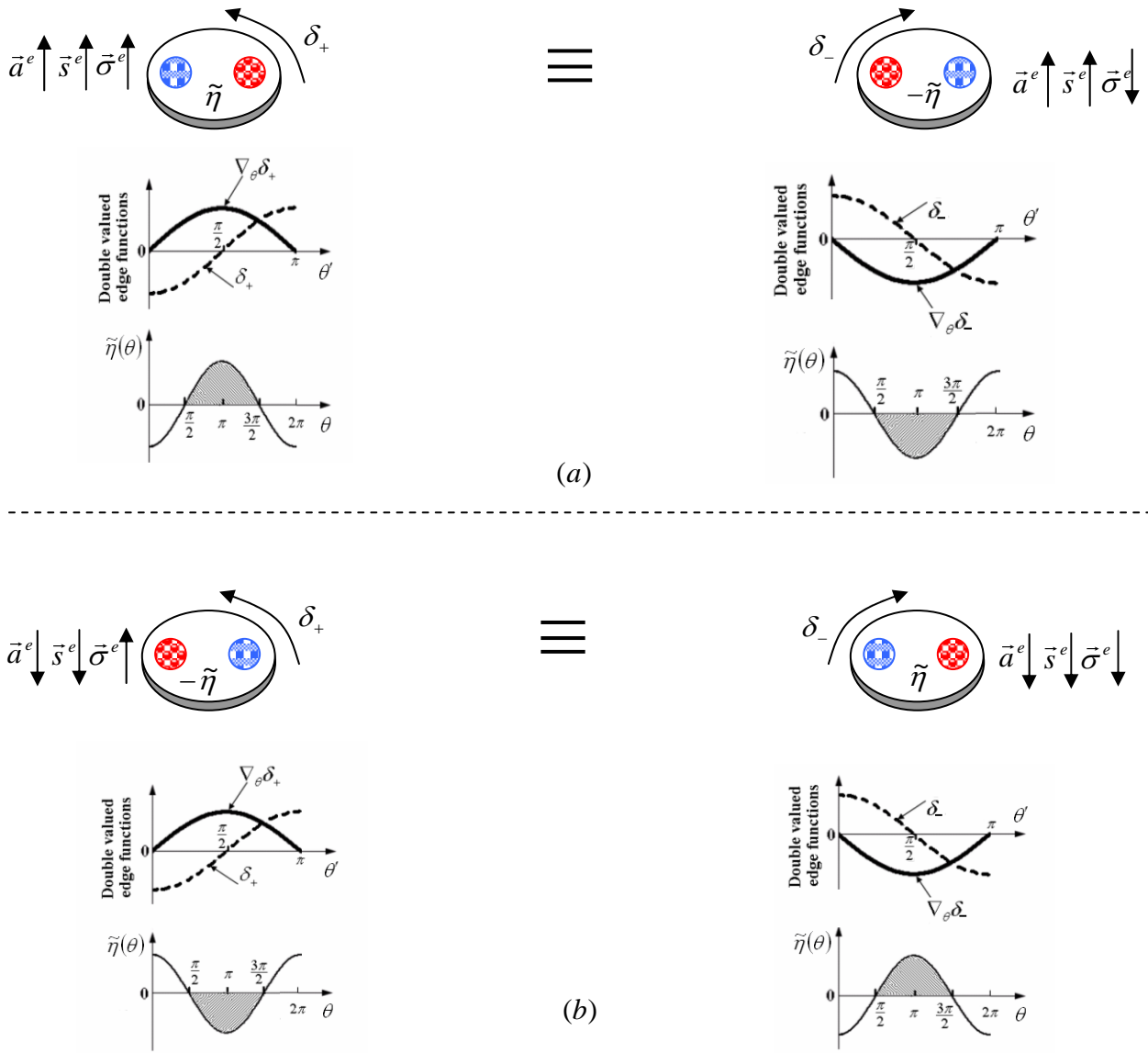

Fig.1. Two cases, (a) and (b), of identical MDM ferrite disks with different distributions of membrane functions $\tilde{\eta}$ and edge functions $\delta$.

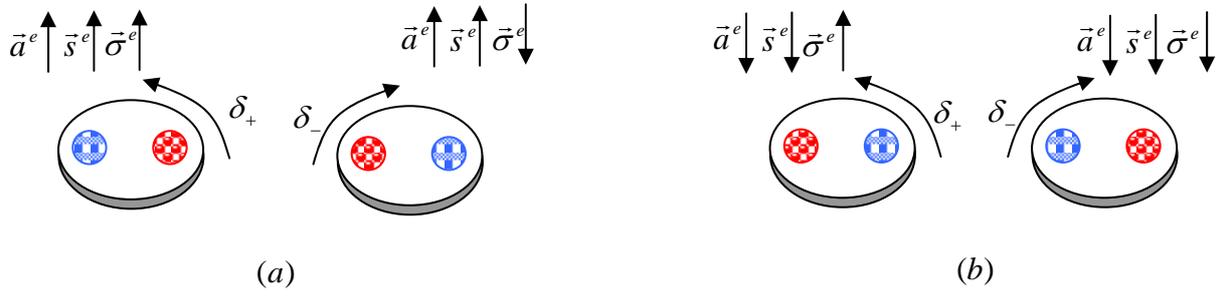

Fig. 2. Two cases of coupled MDM ferrite disks with $\tilde{\eta}^{(S)}$ solutions.

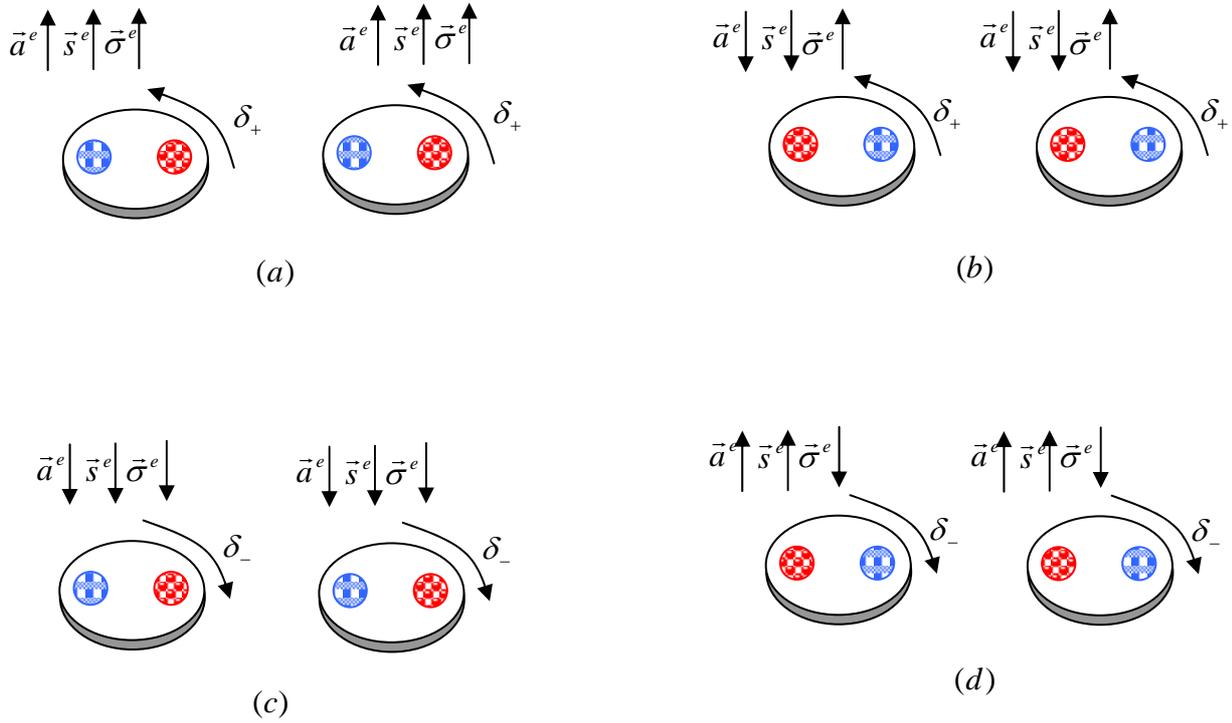

Fig. 3. Four cases of coupled MDM ferrite disks with $\tilde{\eta}^{(A)}$ solutions.